\documentclass[12pt]{article} 
\usepackage{geometry}
\usepackage{caption}
\usepackage{array}

\usepackage{chngcntr}
\counterwithout{equation}{section}

\newcommand{\Team}{2125620} % Replace 1111111 with your Contest Team Control Number
\usepackage{wrapfig}

\usepackage{csvsimple}%for importing the species data from csv
\usepackage{enumitem}%for making lists denser
\usepackage{amsmath,amssymb,amsthm,amsfonts}

\usepackage{hyperref}
\hypersetup{
    colorlinks = true,
    citecolor=black,
    filecolor=black,
    linkcolor=black,
    urlcolor=blue,
    }
\usepackage{lastpage}

\usepackage{booktabs} 

\usepackage{siunitx}

\usepackage{adjustbox}

\usepackage{xcolor}
\usepackage{fancyhdr}
\usepackage{graphicx} % load after xcolor

\allowdisplaybreaks

\usepackage{changepage}

\title{Forays into Fungal Fighting and Mycological Moisture Modeling}
\author{John Blackwelder, Steven DiSilvio, Anthony Ozerov}
\date{February 8, 2021}
\begin{document}

\graphicspath{{.}}  % Place your graphic files in the same directory as your main document
\DeclareGraphicsExtensions{.pdf, .jpg, .tif, .png}
%%%%%%%%%%% Begin Summary %%%%%%%%%%%

\maketitle

\begin{abstract}

As the impending consequences of climate change loom over the Earth, it has become vital for researchers to understand the role that microorganisms play in this process. In this paper, we examine how certain environmental factors, including moisture levels and temperature, affect the expression of certain fungal characteristics on a microscale, and how these in turn affect fungal biodiversity and ecosystem decomposition rates over time. 

We present a differential equation model to understand how the distribution of different fungal isolates depends on regional moisture levels. We introduce both slow and sudden variations into the environment in order to represent the various ways in which climate change will impact fungal ecosystems. This model demonstrates that increased variability in moisture (both short-term and long-term) increases biodiversity and that fungal populations will shift towards more stress-tolerant fungi as aridity increases. The model further suggests the lack of any direct link between biodiversity and decomposition rates.

To better describe fungal competition with respect to space, we develop a local agent-based model (ABM) that focuses on how fungi compete in a 2-dimensional environment. In contrast with the differential equation model, our ABM focuses on individuals, meaning we can track individual fungi and the results of their interactions.  Most notably, our ABM features a more accurate spatial combat system, which allows us to precisely discern the influence that fungal interactions have on the environment. This model corroborates the results of the Differential Equation Model and further suggests that moisture, through its link with temperature and effects on fungal population, also plays a strong role in determining fungal decomposition rates.

When viewed in conjunction, the models suggest that climate change, a phenomenon that portends increasing variability in regional conditions as well as higher average temperatures worldwide, will lead to an increase in both average wood decomposition rates and, independently, fungal biodiversity.

\end{abstract}
\newpage

%%%%%%%%%%% End Summary %%%%%%%%%%%

%%%%%%%%%%%%%%%%%%%%%%%%%%%%%%

% Uncomment the next line to generate a Table of Contents
% change this to {1} if you only want the section titles to appear in the Table of Contents. It is recommended not have subsubsections in the Contents - it become too long.

\tableofcontents 
\newpage

%%%%%%%%%%%%%%%%%%%%%%%%%%%%%%

\section{Introduction}

The term ``fungus'' refers to a large and highly diverse group of organisms \cite{fungi}. Existing research focuses on the relationship between several properties of fungi:
\begin{itemize}
\item Hyphal extension rate: the rate at which a fungus' hyphae, filaments in its mycelium, grow. This controls the rate at which fungi grow.
\item Moisture niche: defined by an optimal moisture level (where the hyphal extension rate is maximized) and the \emph{niche width} (the width of the range of moisture levels at which the hyphal extension rate is at least half of its maximum) \cite{trait-based}. Moisture levels are defined in terms of ``water potential'', which is measured in the unit -MPa, where a high -MPa value corresponds with drier conditions.
\item Competitiveness: how well a fungus competes against other fungi. This depends on intrinsic properties of a fungus, such as the density of its hyphae, as well as its tolerance of its current conditions \cite{untangling}. Competition occurs through a variety of mechanisms, including resource starvation, parasitism, and hyphal overgrowth \cite{combat}.
\item Rate of decomposition: how much material a fungus decomposes in a given amount of time \cite{trait-based}.
\end{itemize}
Fungi with higher moisture tolerances generally have lower decomposition rates, and both more competitive and faster-growing fungi have a higher decomposition rate \cite{trait-based}. It remains to be seen, however, how these factors affect the behavior of ecosystems containing multiple species of fungi interacting with each other, especially under varying moisture conditions.

\section{Problem Statement and Assumptions}
We study how fungi interact with each other and with their environment under a variety of moisture conditions. We model this in a small system containing real fungal isolates (particular genetic instances of species), each with its own hyphal extension rate, competitive rank, moisture niche, and decomposition rate as listed in Table \ref{table:fungi}. As fungi are a very varied group, we specifically focus on the the division Basidiomycota, members of which generally decompose plant materials (especially wood) and use hyphae for vegetative growth \cite{basidiomycota}.  Additionally, we make the following assumptions:
\begin{itemize}
    \item Fungi occupy 2-dimensional space. This simplifies the models, and, after all, much fungal research takes place in effectively 2 dimensional areas, \emph{e.g.} Petri dishes \cite{diversity_begets_diversity}.
    \item Fungi grow from the edges of their area, based on their hyphal extension rate. This homogenizes the various life-cycles of the different species of fungi with respect to spores and fruiting bodies, and existing research on fungi (including those spread over a large area) uses similar techniques \cite{bigboifungus}.
    \item Fungi are circular. Since a fungus starts from a spore, effectively a small circle, and grows from the edges of its area, we can reasonably assume that fungi naturally take a circular shape. Additionally, the tendency for fungi to form circular patterns in nature is well-documented.
    \item Fungal competitiveness in a real combat scenario depends on tolerance of the current moisture conditions. ``Stressful conditions reduce the survival and combative abilities of the most combative species, with negligible effects on the stress-tolerant species'' \cite{untangling}.
    \item Fungal decomposition rate depends solely on the temperature and on the fungus' set decomposition rate at that temperature. This decision stems from fungal experimental data taken at various temperatures \cite{trait-based}.
    \item A fungus' ideal competitiveness rating is an intrinsic property, and we do not take into account historical results of specific pairwise matchups. 
    % A fungus' ideal competitiveness is an intrinsic property of it, and not of it paired with another specific fungus. This simplifies ranking fungi into a simple
    We implement a transitive Elo rating system rather than a pairwise rock-paper-scissors type system. Although intransitive competition is important to explain the behavior of systems with minimal niche differentiation \cite{diversity_begets_diversity}, our modeled systems have significant niche differentiation.
    \item An arid climate corresponds to the BWh Köppen-Geiger climate classification; semi-arid to BSk; temperate to Cfa; arboreal to Dfb; and tropical to Af \cite{koppen}.
\end{itemize}

\section{Modeling Methodology}

% -We only consider multicellular fungi that grow via the production of hyphae, and in particular we look at the 34 species that appear in Table \ref{table:fungi}.
% -Our model allows hyphae to intermingle to some degree, i.e. for fungi to overlap to some degree

\subsection{Differential Equation Model}

\paragraph{Motivation}
Differential equation models are commonly used to model decomposition, describing the biomass of decomposable material and the biomass of the decomposers. Yet under our two-dimensional assumptions, such a model would be inappropriate. Grounding our model in work on predator-prey systems with competition between predators \cite{predator-prey}, we thus propose a differential equation model based on the areas of fungi in a small system. 

\begin{table}[!ht]
\centering
\caption{Differential equation model variables and constants.}
\begin{tabular}[t]{cp{1.8in}lcp{1.9in}}
\hline
Variable &\multicolumn{3}{l}{Description}&Units\\
\hline
$v$&\multicolumn{3}{l}{Dead vegetation area}&m$^2$\\
$d_{r,i}$&\multicolumn{3}{l}{Real decomposition by isolate $i$}&m$^{2}$day$^{-1}$\\
$r_i$&\multicolumn{3}{l}{Radius of isolate $i$}&m\\
$A_i$&\multicolumn{3}{l}{Area of isolate $i$ }&m$^2$\\
$c_i$&\multicolumn{3}{l}{Competition factor for isolate $i$} & -\\
$s_i$&\multicolumn{3}{l}{Moisture suitability factor for isolate $i$ }& -\\
$P_{i,j}$&\multicolumn{3}{l}{Probability of isolate $i$ winning against isolate $j$ }& -\\
$m$&\multicolumn{3}{l}{Current moisture level} &-MPa\\
\hline
Constant &Description&Value&Units&Source/rationale\\
\hline
$l$&Vegetation death rate&$2.7\times10^{-4}$&m$^2$day$^{-1}$&Chosen so that 0.1m$^2$ of vegetation becomes dead vegetation per year.\\
$R_i$&Competitive rank for isolate $i$ &[Table \ref{table:fungi}]&-&Values obtained from \cite{consistent_tradeoffs}.\\
$E_i$&Elo rating for isolate $i$ &[various]&-&Obtained by scaling rank from \cite{consistent_tradeoffs}.\\
$\alpha$&Elo rating scale factor&10&-&Chosen to be low due to the indirect competition in this abstracted model.\\
$\beta$&Minimum moisture suitability factor&0.1&-&Chosen so that fungi far outside their moisture niche may still respond to competition.\\
$d_i$&Maximum decomposition rate for isolate $i$ & [Table \ref{table:fungi}] & m$^{-2}$ day$^{-1}$ & Values obtained from \cite{trait-based}.\\
$h_i$&Maximum hyphal extension rate for isolate $i$ & [Table \ref{table:fungi}] & m\;day$^{-1}$ & Values obtained from \cite{consistent_tradeoffs}.\\
$m_{opt}$&Optimal moisture level for isolate $i$ & [Table \ref{table:fungi}] & -MPa & Values obtained from \cite{consistent_tradeoffs}.\\
$w$&Niche width for isolate $i$ & [Table \ref{table:fungi}] & -MPa & Values obtained from \cite{consistent_tradeoffs}.\\
$I$ & Number of isolates & 34 & - & The number of isolates for which we have cross-referenced data.\\\hline
\end{tabular}

\label{table:DEt}
\end{table}

\paragraph{Dead vegetation}
Dead vegetation, $v$, is described in terms of its area, in m$^2$. It is removed from the system by fungi, and added to the system when living vegetation dies:
\begin{equation*}\label{eq:de_v}
\frac{dv}{dt}=l-\sum_{i=1}^I d_{r,i},
\end{equation*}
where $I$ is the number of fungal isolates in the system, $d_{r,i}$ is the amount of decomposition done by the $i$th fungus under the current conditions, and $l$ is the rate at which dead vegetation is added to the system, assumed to be constant.

%What does the subscript r represent in the above equation?
%it represents that it is the real rate of decomposition… couldn't think of a better variable name lol

\paragraph{Real decomposition rate}
We have data on what proportion of vegetation each fungal isolate consumes in one day. We assume that this was obtained under ideal conditions, with a fungus completely covering the vegetation. Thus, the ideal rate must be scaled:
\begin{equation*}
d_{r,i}=vA_id_i,
\end{equation*}
where $A_i$ is the area of the $i$th isolate and $d_i$ is the daily decomposition proportion of the $i$th isolate. Values of $d_i$ come from the column of Table \ref{table:fungi} appropriate for the climate the simulation is being run in. In this model, arid and tropical climates are taken to have high average temperatures ($22 \ ^{\circ} C$), temperate and semi-arid climates are taken to have medium temperatures ($16 \ ^{\circ} C$), and arboreal climates are taken to have low temperatures ($10 \ ^{\circ} C$) \cite{koppen}.

\paragraph{Fungus growth}
As we assume that fungal growth depends on the extension of their hyphae, and that fungi are circular, we can understand them as growing out from every point on their circumference. Thus, an isolate's growth can be represented in terms of its radius:
\begin{equation*}
\frac{dr_i}{dt}=
h_i c_i,
\end{equation*}
where $r_i$ is the radius of isolate $i$, $h_i$ is the isolate's hyphal extension rate, and $c_i$ is a factor that modulates fungal growth based on competition and carrying capacity. With this equation, the change in area over time can be computed:

%I replaced the "and" right before $c_i(t)$ with a comma 

\begin{equation*}
\frac{dA_i}{dt}=\frac{dA_i}{dr_i}\times\frac{dr_i}{dt}=2\sqrt{A_i}\sqrt{\pi}h_i c_i.
\end{equation*}
To simulate the stress-tolerance and longevity of fungal spores \cite{spores}, isolate areas are forbidden from reaching 0.

\paragraph{Competition and Carrying Capacity} When two fungi meet, their relative competitive ranks determine the outcome. Spatially, we first consider what proportion of a fungus' circumference is taken up by weaker competitors, even competitors, and stronger competitors. If a part of the circumference is taken up by strong competitors, it will retract; if taken up by even competitors, it will neither expand nor retract; %I would say "it will neither expand nor deny" here
if taken up by an extremely weak competitor, it will expand without being hindered by the competitor. %"it will expand without regard for the weaker fungus

Since the differential equation model does not model literal circles, we adapt this idea directly to fungal area. To justify this, consider a small, circular fungus in a region where about half of the area is taken up by a strong competitor. We would expect that, on average, about half of the circumference of the fungus would be in contact with this competitor. We thus model the net factor affecting the fungus' growth by:
\begin{equation*}
c_i=vs_i-\sum_{j=1}^I 2(1-P_{i,j})A_j,
\end{equation*}
where $s_i$ is a factor that adjusts carrying capacity based on moisture, $P_{i,j}$ is the probability that fungus $i$ wins against fungus $j$, and $A_j$ is the area of fungus $j$. Note that the maximum proportion of the circle's circumference where a fungus can expand is regulated by the vegetation density and moisture suitability; if only half of the area contains dead vegetation, but the climate is ideal, a fungus can only expand from around half of its circumference. When $v=1$ and there is no competition, we see that $c_i$ will be directly related to moisture suitability, in accordance with evidence \cite{trait-based}. Note also that, when a fungus encounters a weak competitor, \emph{i.e.} $P_{i,j}=1$, its growth is completely uninhibited. This model also accounts for a fungus competing with itself for area, in which case (as we shall see), $i=j$ and $P_{i,j}=0.5$.

\paragraph{Moisture suitability} One of the key factors in the model is the suitability of a fungus' soil moisture environment. A particular fungus' moisture suitability depends on 3 factors: the moisture level at which its hyphal extension rate is at maximum, the range of moisture levels at which its extension rate is at least half of the maximum (also called the ``niche width''), and the current moisture level of the system. We assume that the niche is centered on the optimal moisture, that moisture suitability follows a downwards-opening quadratic with a maximum of 1, and that the minimum is set to some value. These constraints create the following model:
\begin{equation}\label{eq:suitability}
s_i=\max\left(1-\frac{2(m(t)-m_{opt,i})^2}{w_i^2},\beta\right),
\end{equation}
where $m(t)$ is the current moisture level in the system, $m_{opt,i}$ is the optimal moisture level for fungus $i$, $w_i$ is the moisture niche width for fungus $i$, and $\beta$ is the minimum factor. This minimum allows a fungus to respond to changes in competition even when moisture levels are far outside of its niche.

\paragraph{Win Probabilities}
We have data on the competitive ranks of the 34 fungal isolates in the form of scaled Elo ratings \cite{consistent_tradeoffs}. We rescale these ranks back to Elo ratings based on moisture suitability and a parameter that affects the certainty of victory/defeat:

\begin{equation*}
E_i=\alpha R_i s_i,
\end{equation*}
where $E_i$ is the Elo rating of the $i$th isolate, $\alpha$ is the scaling parameter, $R_i$ is the competitive rank of the $i$th isolate under standard conditions, and $s_i$ is the same moisture parameter for isolate $i$ that affects hyphal extension rate. Given that the original competitive ranks were determined under ideal conditions, that environmental stresses decrease competitiveness, and that it is reasonable to assume that stresses which decrease a fungus' ability to grow also decrease competitiveness, adjusting by $s_i$ is justified. Note that $\alpha$ is chosen to be low, as the competition in this model---with every fungus in contact with every other fungus---is less direct than in a Petri dish, and as competition is straining for any fungus \cite{combat}.

Using these Elo ratings, we can compute every element of a matrix $\mathbf{P}$ of win probabilities by the following formula \cite{elo}:
\begin{equation}\label{eq:probability}
P_{i,j}=\dfrac{1}{1+10^\frac{E_j-E_i}{400}},
\end{equation}
where $P_{i,j}$ is the probability that the isolate $i$ wins against isolate $j$, $E_j$ is the Elo rating of isolate $j$, and $E_i$ is the Elo rating of isolate $i$. Note that, when $j=i$, necessarily $E_i=E_j$ and the win probability is $0.5$. Competition between a fungus and itself is effectively competition between evenly-matched fungi.

\subsection{Agent-Based Model}

\begin{wrapfigure}{R}{0.4\textwidth}
\centering
\includegraphics[width=0.4\textwidth]{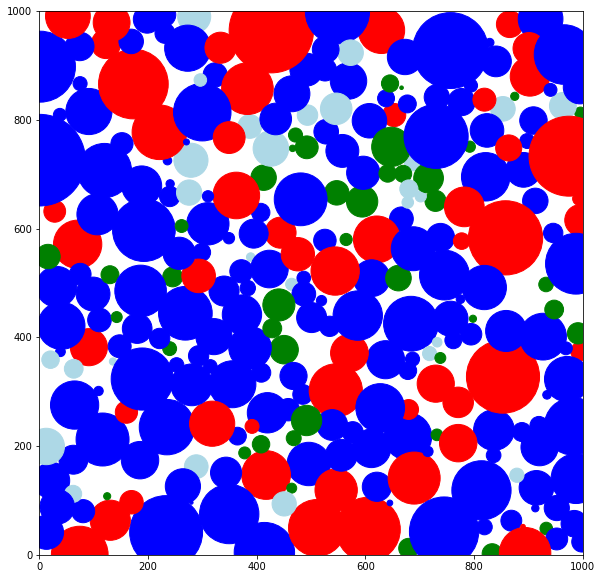}
\caption{Visual representation of spatial fungal distribution in our Agent-Based Model, with each color representing a different isolate group.}\label{fig:splatter1}\vspace{-2.5em}
\end{wrapfigure}  

\paragraph{Motivation}
There is precedent for using an agent-based model (ABM) to represent fungal growth \cite{THOMAS2020100963}. In representing space and individual fungi, this method allows us to more accurately represent the dynamics of a variety of fungal species via the representation of specific traits of interest, namely moisture tolerance and hyphal extension rate, and to provide a more accurate depiction of fungal antagonistic behavior.  Combining these factors with climate data, we may study the population dynamics of the fungi as well as overall levels of decomposition within the environment over time. This model fills gaps in our differential equation model, which does not incorporate space.

\paragraph{Isolate Groups}
Our model exclusively features individual fungi descending from one of the 34 fungal isolates appearing in Table \ref{table:fungi}. However, due to our interest in how fungal moisture tolerance and hyphal growth rate impact overall rates of decomposition in the environment, we divide these 34 isolates into four groups based on these two factors. The groups are designated 1-4 according to the following table:

\def\mca#1{\multicolumn{1}{c}{#1}}
\def\mcb#1{\multicolumn{1}{c|}{#1}}

\begin{tabular}{c|c|c|c|}
  \mca{}  & \mca{\small{Low tolerance}} & \mca{\small{High tolerance}}\\\cline{2-3}
  \mcb{\small{Low extension rate}}   & 1    & 2\\\cline{2-3}
  \mcb{\small{High extension rate}}   & 3    & 4\\\cline{2-3}
\end{tabular}
\\

\iffalse
\newenvironment{myindentpar}[1]%
  {\begin{list}{}%
          {\setlength{\leftmargin}{#1}}%
          \item[]%
  }
  {\end{list}}
  
\begin{myindentpar}{1cm}
\textbf{Group 1}: 
\emph{A. tabescens$_S$,
H. crustosa,
M. meridionalis,
P. robiniae$_S$,
P. robiniae$_N$,
P. hartigii}

\textbf{Group 2}:
\emph{A. sinapina,
A. gallica$_8$,
A. gallica$_2$,
A. gallica$_7$,
A. gallica$_5$,
A. gallica$_1$,
A. tabescens$_N$,
A. gallica$_3$,
A. gallica$_6$,
A. gallica$_4$,
S. commune$_S$,
X. subpileatus}

\textbf{Group 3}:
\emph{T. chioneus,
P. gilvus,
P. pendulus,
H. setigerum$_N$,
S. commune$_N$,
H. setigerum$_S$,
F. fomentarius,
P. sanguineus,
L. confericola,
L. crinitus,
P. flavidoalba$_N$}

\textbf{Group 4}:
\emph{Phlebia acerina$_S$,
P. acerina$_N$,
M. tremellosus,
M. tremullosus,
P. flavidoalba$_S$}

\end{myindentpar}
\fi
\noindent We use the values 3.51 mm/day and 1.56 MPa to divide the groups according to hyphal growth rate and moisture tolerance respectively. Group assignments are shown in Table \ref{table:fungi}.

\paragraph{Model Initialization}
We situate the model within a one square meter, two dimensional region meant to represent the surface of a large decaying log. We believe that this size balances precision with macro scalability, 
% as we expect these results to apply to  
compact enough to accurately assess interactions on a micro scale while still  
providing a wider sense of fungal hierarchies and decomposition rates throughout an entire ecoregion. 
Additionally, the size of the log ensures we don't need to constantly populate the model with dead wood, as we will have more than enough time to make meaningful observations before a log of this magnitude finishes decaying, and we also won't need to model the disappearance of the wood itself. This is a notable simplification when compared to the differential equation model, but it allows us to focus more on the dynamics of fungal interaction, an important aspect of the differential equation model that we wish to corroborate. We typically start with approximately 1000 fungi such that each isolate is represented equally and vary the following parameters between runs:
\begin{itemize}

\item \textbf{Climate}: Using localized historical weather and moisture data \cite{moisture}, we are able to accurately represent various types of climates during our run. In particular, we focus on arid, semi-arid, temperate, arboreal, and tropical rainforest regions. %On some runs of the model we implement sharp deviations from the historical moisture data in order to simulate the consequences should such an event actually occur.
\item \textbf{Combat Parameters:} One of our motives for implementing an ABM was the potential for recreating realistic fungal interaction. As we mention in the introduction, the interaction between fungi is incredibly complex and comes in many different forms. We do a brief study on our combat system in order to gauge how certain assumptions impact the environment. Key to our modeling of fungal competition is that competition detriments both the stronger and weaker opponents, with no fungus actively engaged in competitive interactions able to expend energy towards hyphal growth.  
% Additionally, 
\end{itemize}

%\paragraph{Assumptions}

\begin{table}[!ht]
\centering
\caption{Agent-based model variables and constants}
\begin{tabular}[t]{cp{2in}ccp{1.8in}}
\hline
Variable &\multicolumn{3}{l}{Description}&Units\\
\hline
$w_f$&\multicolumn{3}{l}{Wood decomposed by fungus $f$ on a given day } &mm$^{2}$\\
$r_f$&\multicolumn{3}{l}{Radius of $f$}& mm\\
$A_f$&\multicolumn{3}{l}{Area of $f$}&mm$^2$\\
$s_f$&\multicolumn{3}{l}{Moisture suitability factor of $f$}&-\\  
$P_{f,g}$&\multicolumn{3}{l}{Punishment score incurred on fungus $f$ by fungus $g$}&-\\
\hline
Constant &Description&Value&Units&Source/rationale\\
\hline
$d_f$&Maximum decomposition proportion for fungus $f$ & [various] & day$^{-1}$ & Piecewise continuous function derived from linearly connecting values from \cite{trait-based}. \\
$R_f$&Competitive rank for fungus $f$ &[Table \ref{table:fungi}]&-&Values obtained from \cite{consistent_tradeoffs}.\\
$E_f$&Elo rating for fungus $f$ &[various]&-&Obtained by scaling rank from
\cite{consistent_tradeoffs}.\\
$p$&Constant scaling the impact of fungal interaction & varies & mm & Chosen to create a reasonable balance between fungal competitive capabilities and other traits \\
$h_f$&Maximum hyphal extension rate for fungus $f$ & [Table \ref{table:fungi}] & mm\;day$^{-1}$ & Values obtained from \cite{consistent_tradeoffs}.\\
$\alpha$ & Elo rating scale factor & 1000 & - & Chosen to be high to represent direct competition between fungi.
\end{tabular}
\label{table:ABMt}
\end{table}

%\subsubsection{Environmental Mechanisms}
%Maybe don't need this one?

\subsubsection*{Fungal Characteristics and Behavior}
Every day, fungi decompose wood, expand their radii, and may enter into conflict with neighboring fungi. We describe these behaviors in detail.

\paragraph{Area}
Keeping in mind the assumption that our fungi are circles, we calculate the area of a fungus $f$, $A_f$, with the following equation:
\[A_f = \pi r_f^2, \]
where $r_f$ is the fungal radius.

\paragraph{Decomposition}
% The entire area that a fungus occupies must contain wood in order for the fungus to engage in decomposition. If this condition is not met, then a fungus will not absorb nutrients and consequently will not grow. 
Assuming that all instances of a given fungal isolate share the same innate attributes, we first find $d_f$, the daily base decomposition rate constant for fungus $f$, by linearly connecting the points obtained from \cite{trait-based}. We set 0 and 50 mm as minimum and maximum growth rates respectively in order to insure that this method doesn't result in any unreasonable values. Then, we use the following equation to determine $w_f$, the amount of wood that the fungus decomposes:
\[w_f = d_f A_f.\]

\paragraph{Growth}
We calculate the moisture suitability value, $s_f$, using Equation \eqref{eq:suitability} in the same way as in the differential equation model. 
%Using moisture data from (CITE), we run the same equations as in the differential equations model (BETTER WAY TO REFERENCE THESE?) in order to come up with a moisture suitability value. 
Each day, the radius of a given fungus increases by $s_fh_f$, assuming that it did not engage in combat with any surrounding fungi.

\paragraph{Combat} 
Combat occurs in the event two individual fungi attempt to expand to, or already occupy, the same region. For every such instance, a fungus will enter into combat, and hence fungi can combat multiple times per day.  The factors that determine the outcomes of fungal interactions in nature are numerous and well-documented, but we base our model's combat mechanism on the observation that all fungi involved, even those that end up dominating neighboring fungi, often suffer inhibited or even negative growth throughout the duration of their interaction \cite{combat}. We purposefully differentiate the ABM combat mechanism from that of the differential equation model in order to better understand the relevance of this observation. Another distinction is that we are applying it to well-defined, individual agents instead of an abstracted conception of area.

First, we calculate the Elo rating for a fungus $f$, $E_f$, as follows:
\[E_f=\alpha R_f s_f, \]
where $\alpha$ is the Elo rating scale factor, and $R_f$ is fungus' competitive rank. We then calculate $P_{f,g}$, the punishment score for $f$ with respect to $g$, via an alteration of Equation \eqref{eq:probability}:
\begin{equation*}
P_{f,g}=\dfrac{1}{1+10^\frac{E_f-E_g}{400}}.
\end{equation*}
Note that here the more competitively able a fungus is against its opponent the lower its resulting punishment score.

The total radial distance fungus f loses in combat against fungus g is then given by:
\[P_{f,g} \times p \, \, ,\]
with $p$ being  the punishment constant, set to represent the maximum radial distance a fungus can lose from any given interaction. We then prevent both $f$ and $g$ from growing later in the day in order to simulate the inhibitory effects of antagonistic interaction, such as through hyphal interference or parasitism. A fungus dies when its radius reaches zero according to the above equation.

Note that we set $p$ to equal 1 for all runs of the model unless specified otherwise.

% Although the ABM takes advantage of the same ELO ranking system used in our previous model, it applies it differently. When two fungi, $f$ and $g$, enter a mutual state of combat, we use the ELO formula described previously in order to calculate the relative damage sustained by each of them. The fact that each individual fungus derives its combat effectiveness rating solely from its source isolate is allows us to do this.  

%\paragraph{Note regarding reproduction}
% The reader may wonder at this point why we did not implement any forms of fungal reproduction or fungal death (independent of combat) in the ABM. Although individual fungal fruit bodies form and die within a matter of weeks, fungal colonies tend to persist for a much greater amount of time. Our model takes advantage of the blurred definition  

\section{Model Results}

\subsection{Differential Equation Model}
We use an Euler approximation to numerically solve the Differential Equation Model in a number of different environments. In all runs, dead vegetation is initialized at a density of 1. Fungal areas are initialized at $\frac{1}{I}$, such that their areas are equal and sum to 1.
\subsubsection{Sample solution}

\begin{figure}[!h]
 \centering
  \includegraphics[width=0.8\textwidth]{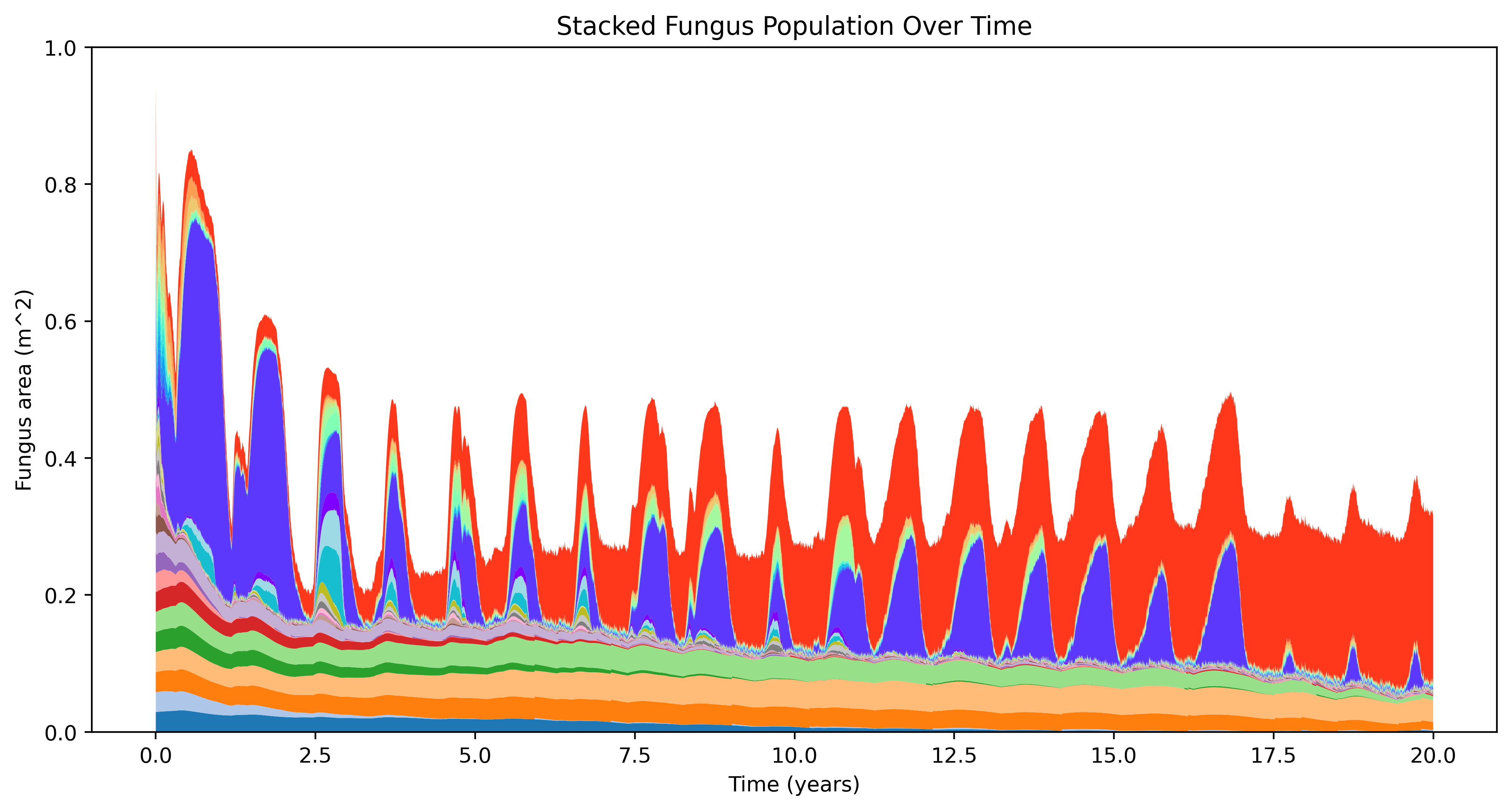}
  \includegraphics[width=0.8\textwidth]{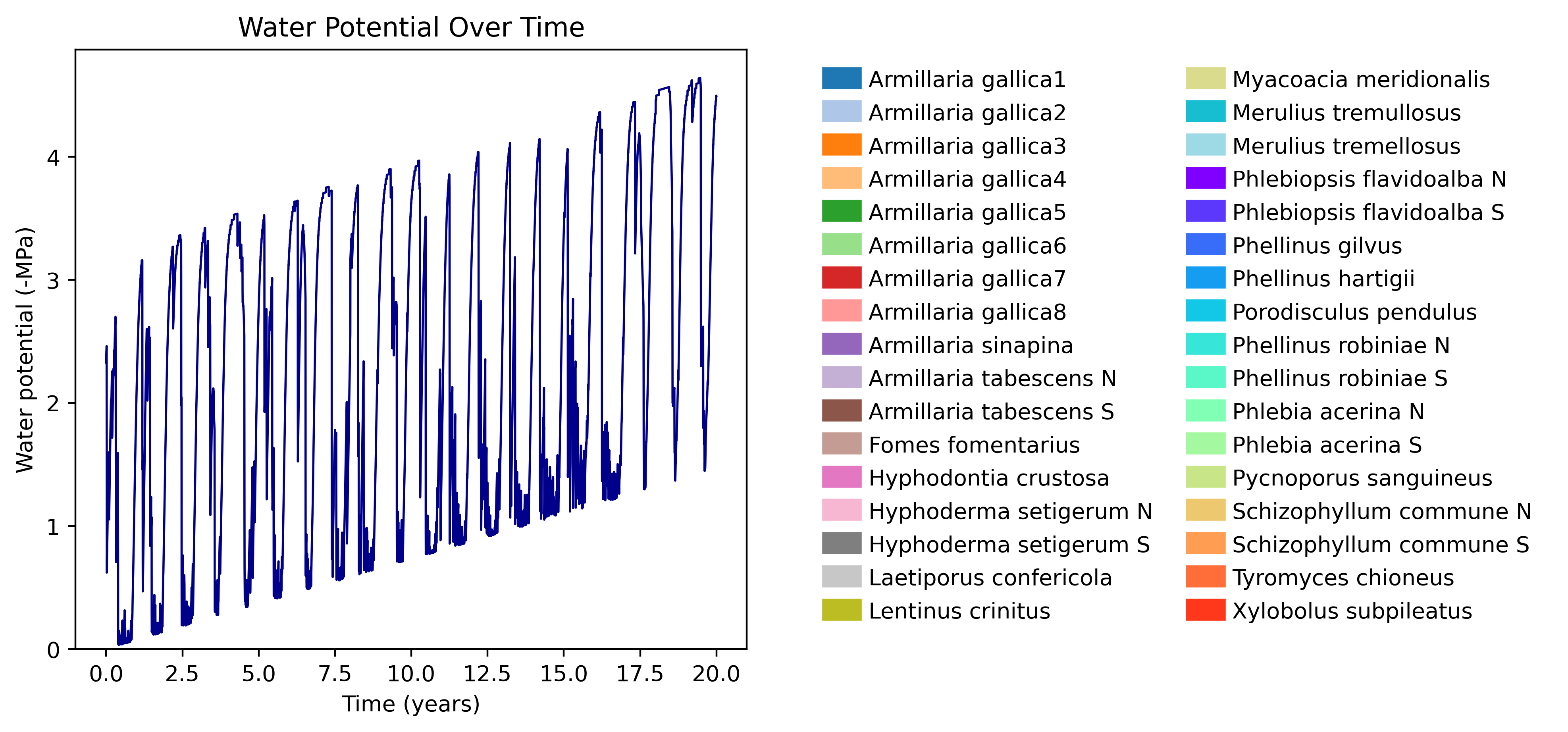}
  \caption{Numerical solution to the Differential Equation Model for 20 years of run-time in an arid climate. Water potential data are taken from 16.8°E, 13.9°N, and slowly increased over time to simulate increasing dryness. \emph{P. flavidoalba S} initially dominates, but after multiple dryness shocks \emph{X. subpileatus} begins to dominate.}\label{fig:de_full}
 \end{figure}  

We run the model for 20 years in an arid climate in order to convey fungal population dynamics and the progress of the model. The results are illustrated in Figure \ref{fig:de_full}. 

\paragraph{Moisture and competition}
One isolate, \emph{P. flavidoalba S}, begins to dominate within a year. It has the highest rank out of any isolate and a large moisture niche width, giving it a competitive advantage. However, as time continues, the location experiences increased dryness and frequent dryness shocks. This decreases the Elo rating and carrying capacity of \emph{P. flavidoalba S} more so than it does for an isolate like \emph{X. subpileatus}. This is because, as seen in Table \ref{table:fungi}, the niche width for \emph{X. subpileatus} is much wider than that for \emph{P. flavidoalba S}, thus, the moisture shocks affect the latter more, and the former is able to fill the room left in the ecosystem. %Once this room is filled, variation in the total area of the fungi decreases, which also results in decreased variation in the consumption of dead vegetation.

\paragraph{Coexistence under variation}
In the time period from about 7.5 years to 15 years, both isolates are present simultaneously. Evidently, this represents a period of time where moisture levels fluctuate between the regimes where one or the other is dominant, hence creating biodiversity through variation.

\paragraph{Climate shocks}
This example also illustrates the importance of climate shocks in inhibiting dominant isolates. We see that, after every spike in dryness, the area covered by \emph{P. flavidoalba S} decreases, which causes a spike in the areas of isolates like \emph{M. tremullosus}.

\subsubsection{Large-sample statistics}
To get a more broadly-applicable set of results, we run the model for 5 years in 50 different locations each for arid, semi-arid, temperate, arboreal, and tropical climates. Using these data, we examine how climate variability affects biodiversity and biodiversity affects decomposition.
\paragraph{Environment}
For each simulation of each climate type, we select a random point on the Earth of that climate according to data \cite{koppen}. The water potential at that point is approximated from soil water volume data \cite{moisture}.
\paragraph{Biodiversity Measure}
The Simpson index, used to measure biodiversity, is defined by the probability of randomly selecting two individuals and having them be the same \cite{simpson}. Adapting this to fungi, we can describe it as the probability of selecting two points covered by fungi and having them be the same fungus. We can thus describe it by the equation:
\begin{equation*}
    \lambda=\sum_{i=1}^I \left(\frac{A_i}{A_T}\right)^2,
\end{equation*}
where $\lambda$ is the Simpson index, $A_i$ is the area of isolate $i$, and $A_T$ is the total area of the isolates. A lower Simpson index represents greater biodiversity.

\begin{figure}[!h]
 \centering
 \begin{minipage}{0.49\textwidth}
    \centering
    \includegraphics[width=1\textwidth]{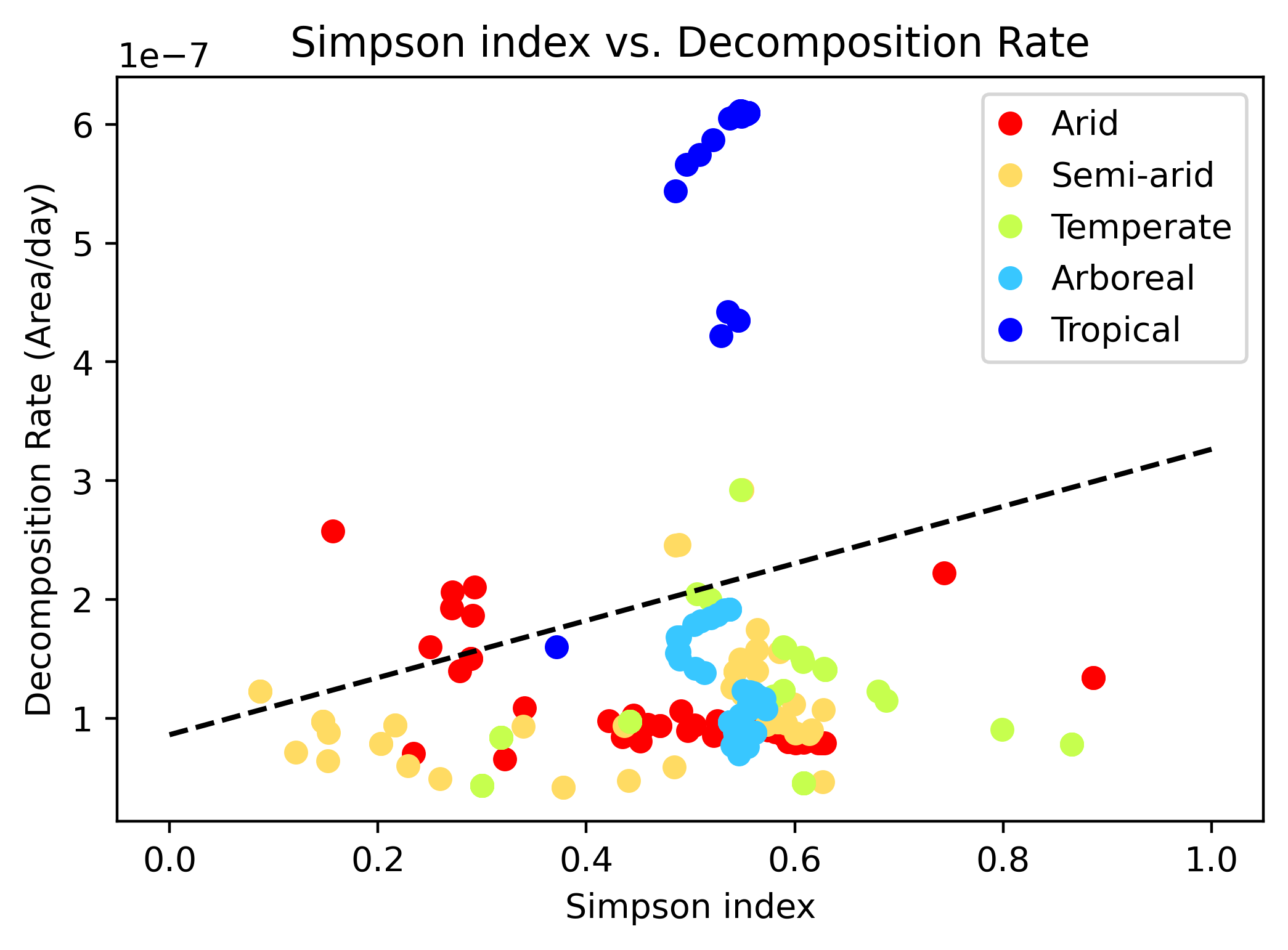}
    \caption{Simpson index vs. Decomposition rate in 250 locations in 5 different climate types. No clear association is evident, especially given the clustering of Tropical points far from the rest. Linear regression slope=$2.4\times 10^{-7}$, intercept=$8.6\times 10^{-8}$, $R^2$=$0.028$.} \label{fig:simpson-decomp}
 \end{minipage}\hfill
 \begin{minipage}{0.49\textwidth}
    \centering
    \includegraphics[width=1\textwidth]{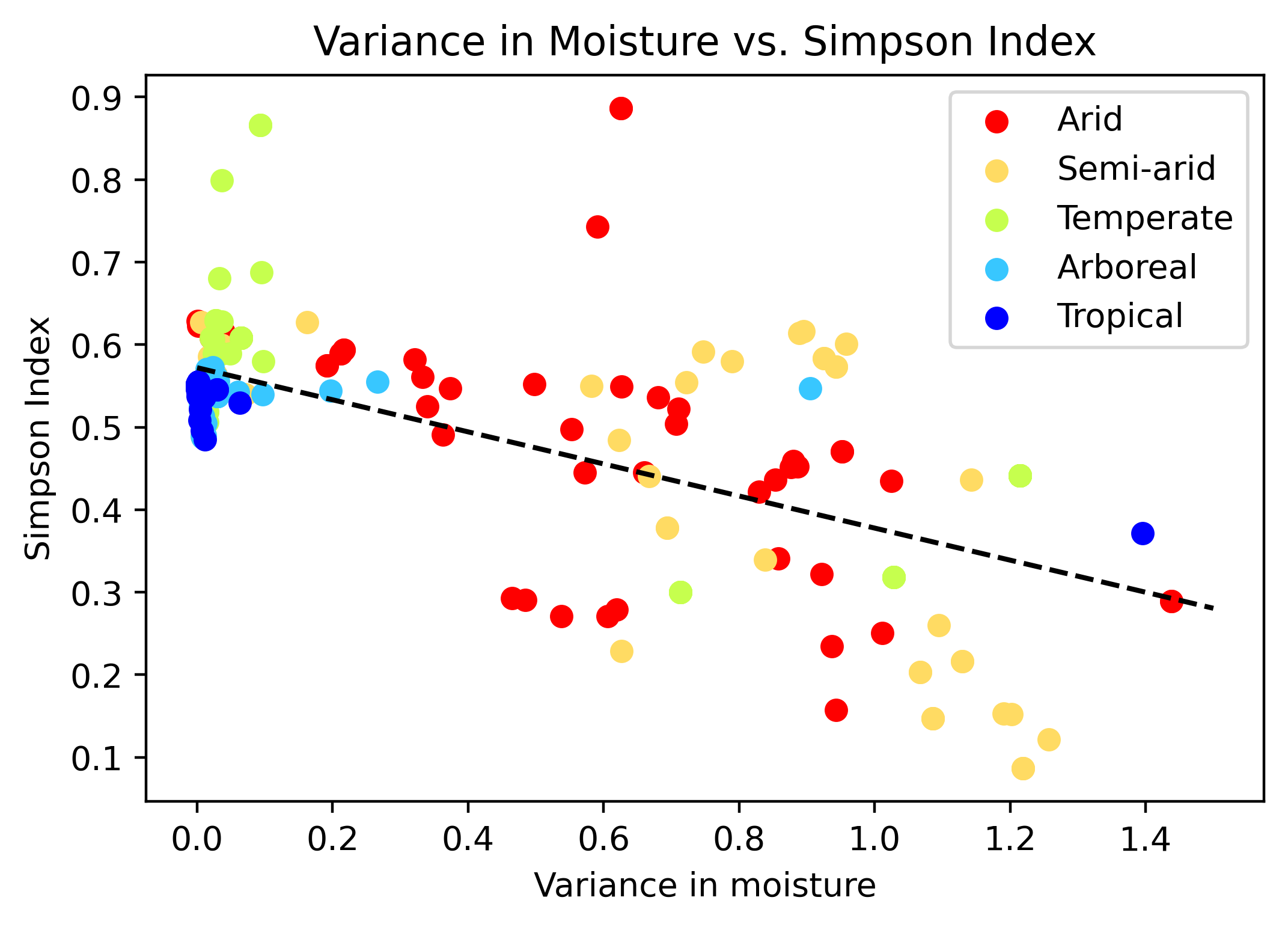}
    \caption{Moisture variance vs. Simpson index in 250 locations in 5 different climate types. A negative association is evident. Linear regression slope=$-0.19$, intercept=$0.57$, $R^2$=$0.42$.}\label{fig:variance-simpson}
 \end{minipage}
 \end{figure}  

\paragraph{Biodiversity and decomposition}
The relationship between biodiversity and decomposition becomes evident when the Simpson index at the end of 5 years is plotted against the decomposition rate. Figure \ref{fig:simpson-decomp} plots this relationship, including the results for each climate. We observe a weak association between the Simpson index and the decomposition rate, with $R^2=0.028$. We are unable to conclude that such a relationship really does exist.

The lack of any relationship between biodiversity and decomposition rate might be because, while highly dominant fungi generally have high decomposition rates \cite{trait-based}, the isolates that end up dominating in our scenario, such as \emph{P. flavidoalba S}, do not have particularly high decomposition rates. However, when isolates are randomly excluded from the model, there is, too, no relationship between the number of isolates included in the model and the decomposition rate after five years. More research is required into this relationship.

\paragraph{Moisture Variation and Biodiversity}
The relationship between variation in moisture levels and biodiversity becomes evident when the variance in moisture across the full 5 years in each location is compared with the Simpson index. Figure \ref{fig:variance-simpson} plots this relationship. We observe a strong negative linear relationship between moisture variation and the Simpson index rate, with $R^2=0.42$. Thus, 42\% of the variance in the Simpson index is explained when the degree of moisture variability is accounted for, suggesting an association between the two factors. We also understand the causal nature of this result: as variation in climate increases, ultra-competitive isolates like \emph{P. flavidoalba S} become less competitive than isolates like \emph{X. subpileatus} which inhabit a much wider moisture niche. These tolerant isolates do not compete as strongly with other tolerant isolates, hence biodiversity increases. Thus, with the observed relationship and the sensible causal mechanism, we conclude that increased variation in moisture causes increased biodiversity.

\begin{figure}[!ht]
 \centering

\includegraphics[width=1\textwidth]{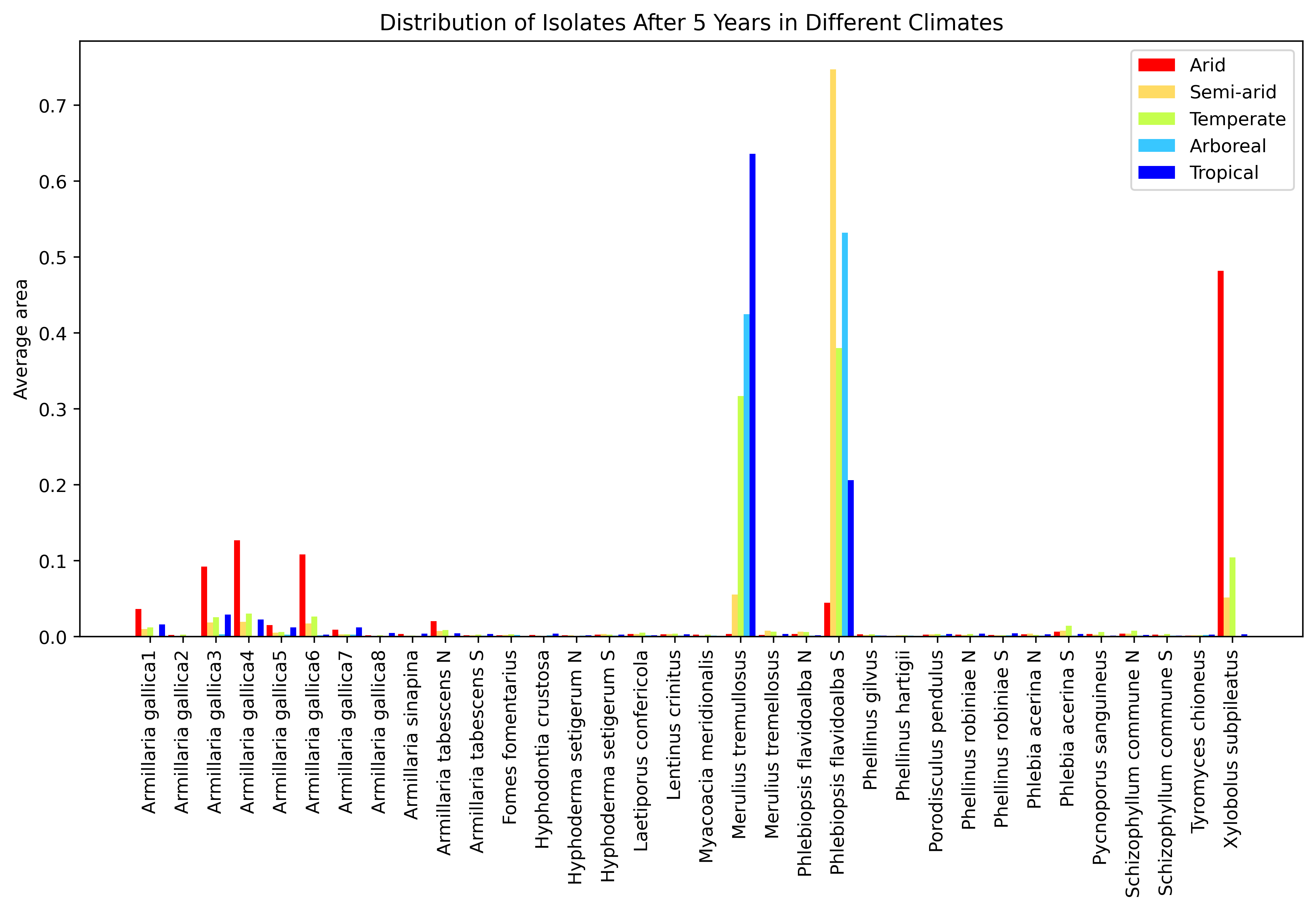}
\caption{The average area (relative to the total area) for each isolate in each climate type. We see that \emph{P. flavidoalba S} dominates semi-arid locations, and that \emph{M. tremullosus} and \emph{P. flavidoalba S} dominate temperate, arboreal, and tropical locations, with the former performing better in tropical locations. In arid locations, \emph{X. subpileatus} dominates, though isolates of the \emph{A. gallica} species are also present.} \label{fig:distribution}
\end{figure}

\paragraph{Isolates likely to survive}
The isolates likely to survive under different conditions can be deduced from the dominant isolates after 5 years. Figure \ref{fig:distribution} displays the average proportion of the fungal population made up by each isolate at the end of 5 years. Evidently, in every climate, only a few isolates dominate. 
\begin{itemize}
    \item In tropical locations, \emph{M. tremullosus} dominates. This is likely because of its high competitive rank, which, when combined with its high optimal moisture level, leads it to generally out-compete other isolates in the moist conditions of said locations. 
    \item In temperate and arboreal climates, \emph{M. tremullosus} and \emph{P. flavidoalba S} together dominate the ecosystem. Both are highly competitive, and variance in moisture is generally low in these regions (as evidenced in Figure \ref{fig:variance-simpson}), allowing both to out-compete other isolates. In Figure \ref{fig:variance-simpson} it is also evident that the temperate regions with high variance in moisture had much more biodiversity, suggesting that these two isolates do not compete as well under those conditions.
    \item In semi-arid regions, \emph{P. flavidoalba S} out-competes all other isolates. Its high competitiveness and relatively high moisture niche width allow it to compete well under moderately variable circumstances.
    \item In arid regions, the dynamics shift and \emph{X. subpileatus} begins to dominate. Its middling competitive rank of 0.49 is compensated by the width of its moisture niche---4.96 MPa---which allows it to compete in extremely dry conditions. Joining \emph{X. subpileatus} are isolates of the \emph{A. gallica} species, especially 3, 4, and 6. These isolates, though they have lower competitive ranks than \emph{X. subpileatus}, have comparably wide moisture niche widths which allow them to survive under dry conditions.
\end{itemize}

%\paragraph{Biodiversity in different climates} ??
%[histogram of different climates and the biodiversity in them]

\subsection{Agent-Based Model}
\subsubsection{Sample result}
\begin{figure}[!ht]
    \centering
    \begin{minipage}{0.49\textwidth}
        \centering
        \includegraphics[width=1\textwidth]{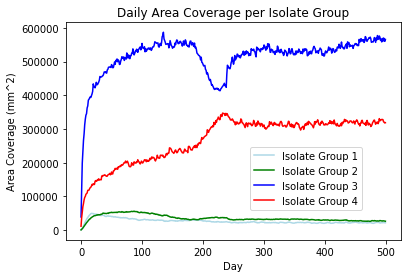}
        \caption{The area that each isolate group covers per day. The groups quickly establish a dominance hierarchy and each of their rates of change begin to level off, but then we see a sharp drop off in group 3 area covered accompanied by an increase in groups 4 and 2.}
        \label{fig:biomass1}
    \end{minipage}\hfill
    \begin{minipage}{0.49\textwidth}
        \includegraphics[width=1\textwidth]{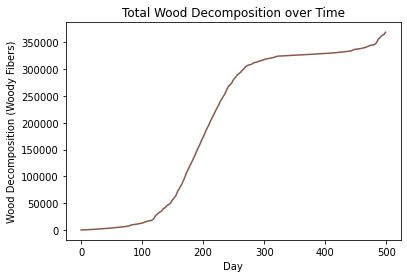}
        \caption{The total decomposition of wood in a semi-arid region. Measured in terms of woody fibers decomposed. Note the fluctuations in decomposition rate; we see a more linear graph in less variable climates.}
        \label{fig:decomp1} 
    \end{minipage}
\end{figure}

\begin{figure}[!ht]
    \centering
    \begin{minipage}{0.49\textwidth}
        \centering
        \includegraphics[width=1\textwidth]{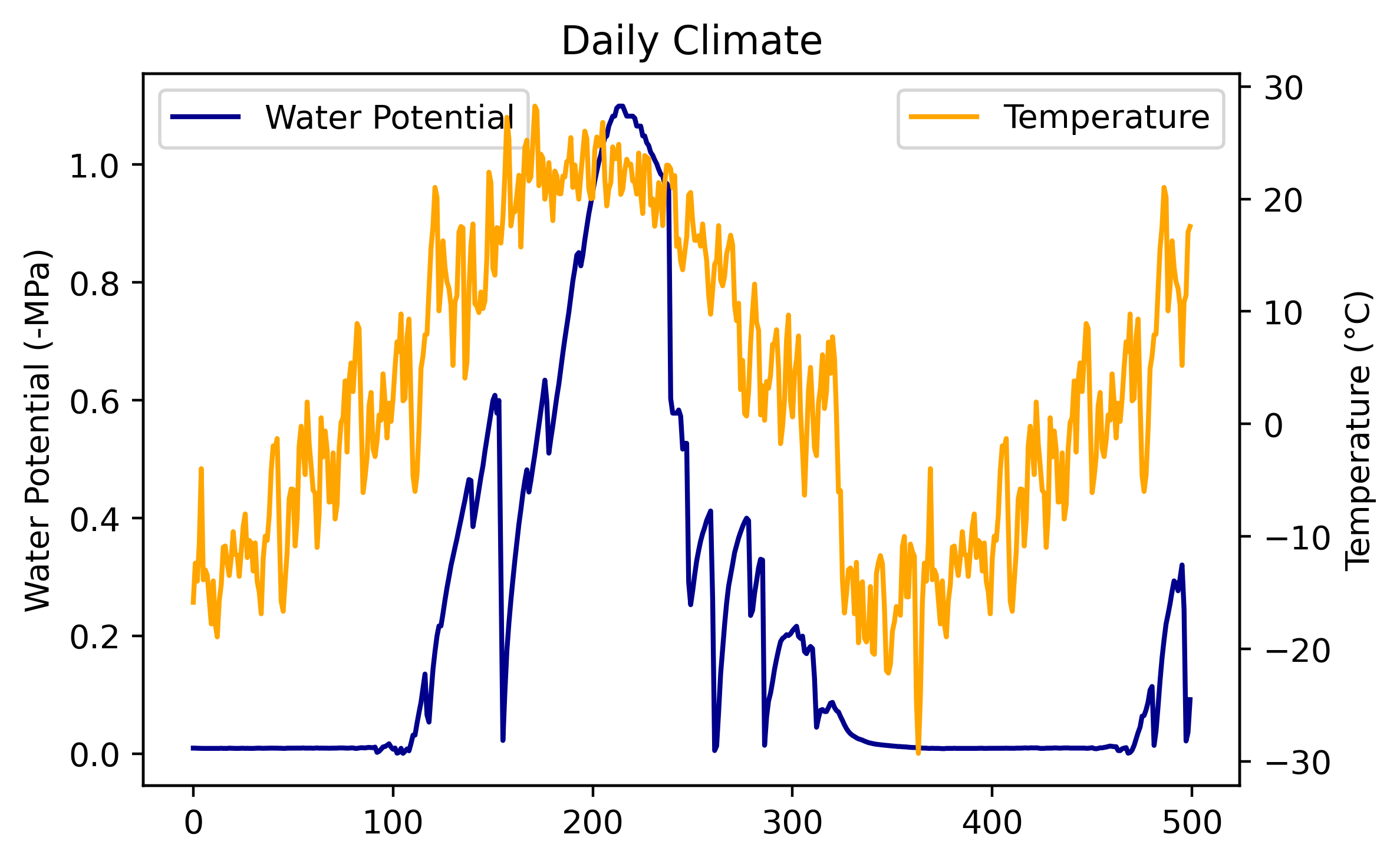}
        \caption{Daily moisture levels, measured in -MPa, alongside daily temperature levels, in $^\circ$C. Notice the sharp decrease in soil saturation beginning around day 100. Dryness reaches a peak simultaneously with temperature.}
        \label{fig:moisture1}
    \end{minipage}\hfill
    \begin{minipage}{0.49\textwidth}
        \includegraphics[width=1\textwidth]{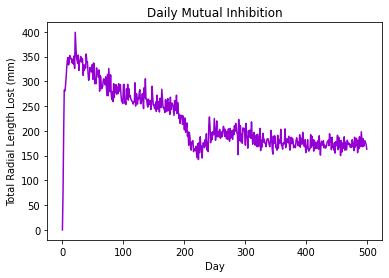}
        \caption{Daily growth inhibition due to fungal interaction, measured in terms of total radius lost. }
        \label{fig:combat1} 
    \end{minipage}
\end{figure}

We present the result of our model in a semi-arid region, initializing it with 1,000 individual fungi, and allowing it to run for 500 days. The results are depicted in Figures \ref{fig:biomass1}, \ref{fig:decomp1}, \ref{fig:moisture1}, and \ref{fig:combat1}.

\paragraph{Species distribution and competition}
The four isolate groups establish a status quo almost immediately, with group 3 dominating in terms of area coverage. However, a sudden decrease in moisture levels occurring from approximately day 150 to day 200 coincides with a sharp drop off in group 3 area coverage and a corresponding increase in the area covered by groups 2 and 4, as they are more tolerant of moisture variation. This corroborates the results of the Differential Equation Model regarding dryness shocks. The measurements of competition in Figure \ref{fig:combat1} also reveal that the moisture spike, in reducing the population of group 3 fungi, caused a general reduction in the level of competition in the system. The increase in the population of group 4 and group 2 fungi can be attributed to this.

\paragraph{Decomposition}
The rate of environmental wood decomposition also increases during the dry period. We find that this is attributable to the increased temperature associated with the period of reduced moisture, as well as the population changes that occur due to the reduced moisture. A similar connection between reduced moisture levels and increased rates of decomposition occurs on approximately day 480, albeit to a much smaller degree. Given that the relative areas don't change much around that day, we attribute the increased decomposition there to increased temperatures.

\subsubsection{General Results}
To obtain more general results from the Agent-Based model, we run it in a variety of different climates. The results of simulations in arid, semi-arid, temperate, arboreal, and tropical climates are displayed in Table \ref{tab:agent_table}. 

\begin{table}[ht]
\centering
\begin{adjustbox}{width=1\textwidth}
\small
\begin{tabular}{rlrrrrrrrr}
  \hline
 Climate & \%SG1 & \%SG2& \%SG3 & \%SG4 & Total Area & Decomposed Wood (mm$^2$)& Alive\\
  \hline
Arid & 2.35 & 16.30 & 24.58 & 56.77 &  877,523 & 1,510,495&184 \\ 
  Semi-Arid & 2.80 & 3.91 & 68.81 & 24.55 &  924,427 & 371,317&380\\ 
  Temperate & 5.64 & 4.28 & 73.86 & 16.21 &  918,818 &701,964&371    \\ 
  Arboreal& 3.58 & 3.15 & 71.13 & 22.14 & 934,267 & 749,129 &324 \\ 
  Rainforest & 3.27 & 4.44 & 69.13 & 23.16 & 890,572 & 1,344,068&327 \\ 
\hline
\end{tabular}
\end{adjustbox}
\caption{Results of the Agent-Based Model for various climates after 500 days of simulation.} \label{tab:agent_table}
\end{table} 

\paragraph{Climate and Distribution}
Isolate Groups 2 and 4, both having high moisture tolerance, perform much better in arid climates, than they do in other climates. Group 3, with low moisture tolerance, performs much better in non-arid climates. This aligns with our expectations and corroborates the results of the Differential Equation Model. It is also notable that, under the climate stresses of arid regions, fewer individuals survive (though, combined, their total area is roughly the same). While this does not necessarily reflect a reduced level of biodiversity, it does capture the fact that the harsh climate conditions reduce the number of individuals able to coexist.

\paragraph{Decomposition}
This model's more complex model of wood decomposition also reveals differences in the amount of decomposed wood in different climates. Table \ref{tab:agent_table} suggests that temperature is the factor most affecting long term wood decomposition. Both arid and rainforest regions, having higher temperatures, have greater overall wood decomposition, even though they differ widely in their moisture levels and isolate distribution.

%\begin{table}[ht]
%\centering
%\begin{adjustbox}{width=1\textwidth}
%\small
%\begin{tabular}{rlrrrrrrrr}
%  \hline
% Ecoregion & SG1 (mm$^2$) & SG2& SG3 & SG4 & TFA & Decomposed Wood& Alive\\
%  \hline
%Arid & 20,629 & 143,052 & 215,715 & 498,127 &  877,523 & 1,510,495&184 \\ 
%  Semi-Arid & 25,911 & 36,149 & 636,068 & 226,928 &  924,427 & 371,317&380\\ 
%  Temperate & 51,869 & 39,323 & 678,625 & 149,000 &  918,818 &701,964&371    \\ 
%  Arboreal& 33,451 & 29,453 & 664,556 & 206,805 & 934,267 & 749,129 &324 \\ 
%  Rainforest & 29,149 & 39,501 & 615,626 & 206,293 & 890,572 & 1,344,068&327 \\ 
%\hline
%\end{tabular}
%\end{adjustbox}
%\caption{Agent based model results for various ecoregions} \label{tab:agent_table}
%\end{table} 

\subsection{Result Synthesis}
\subsubsection{Short-term trends}
\paragraph{Climate shocks}
Our results indicate that climate shocks are important in promoting the biodiversity of a population of fungi. Sudden moisture stress on dominant isolates reduces their area and how successfully they can compete against other isolates. As shown in the Agent-Based Model, this reduces competition and allows other isolates to proliferate, at least until dominance is re-established.

\subsubsection{Long-term trends}
\paragraph{Desertification}
Many regions on the Earth are becoming drier as a result of global warming, particularly semi-arid regions \cite{desertification}. Our Differential Equation Model indicates that as semi-arid regions become arid regions, fungi with wide moisture niches, even if not intrinsically competitive, will come to dominate. They, like the prototypical \emph{X. subpileatus}, will be able to out-compete highly competitive fungi under extremely dry conditions. Even very noncompetitive isolates like isolates 3, 4, and 6 of \emph{A. gallica} will perform well due to their wide moisture niches. The transition from semi-arid to arid will also be associated with increased biodiversity as it will no longer be possible for a single highly competitive isolate such as \emph{P. flavidoalba S} to dominate; competitive isolates, according to the dominance-tolerance tradeoff, are generally less tolerant to moisture stresses \cite{trait-based}. Our Agent-Based Model corroborates this result in showing that arid climates promote stress-tolerant isolates. It further indicates that desertification may result in higher average rates of decomposition worldwide.

\paragraph{Climate variability}
There is evidence that global warming has been increasing the frequency of extreme climate events, including precipitation \cite{ipcc}. Our Differential Equation Model shows a strong negative relationship between variance in moisture and the Simpson index, which corresponds to a strong positive relationship between variance and fungal biodiversity. We thus predict increased fungal biodiversity as a result of this long-term trend. None of our results indicate that biodiversity impacts the overall rate of decomposition, but our Agent-Based Model suggests that climate variability (especially with respect to temperature) causes variation in total decomposition by varying the decomposition rates of individual fungi as well as changing the population distribution.

\section{Model Evaluation}

\subsection{Common Strengths and Limitations}
Our two models, both having some assumptions in common, have a set of common strengths and limitations.
\paragraph{Strengths}
\begin{itemize}
    \item Our models can be directly implemented for any isolates of Basidiomycota for which the requisite data is available.
    \item Using real-world moisture data allows us to realistically incorporate real-life events like dryness shocks without making assumptions about their duration, intensity, or frequency.
\end{itemize}
\paragraph{Limitations}
\begin{itemize}
    \item In only considering Basidiomycota, our models are less applicable to types of fungi with different growth mechanisms.
    \item As fungi are thought of as always being circular, fungal growth does not take into account the different shapes they would take if forced to grow in a particular direction. This makes our models for fungal competition less realistic.
    \item Soil water potential, in MPa, involves much more than simply the amount of water in it, depending on other factors like the mineral contents of the soil. While we are able to capture overall variation, we are unable to capture how the fungi themselves and the decomposition process affect the water potential.
\end{itemize}

\subsection{Differential Equation Model}
\paragraph{Sensitivity Analysis}
We performed a sensitivity analysis on the Elo rating scaling parameter $\alpha$. While levels of biodiversity decreased as $\alpha$, which emphasizes competitive advantage, increased, the overall distribution of isolates remained the same. We also performed a sensitivity analysis on $\beta$, and found that increasing $\beta$ makes the fungi less sensitive to climate shocks. This allows strong competitors to survive under more varied conditions, hence decreasing biodiversity, yet the general trend again remained the same. While adjusting $\alpha$ and $\beta$ does, necessarily, change the outcome of the model, the conclusions drawn are robust.
\paragraph{Strengths}
\begin{itemize}
    \item Our results about stress-tolerant fungi dominating in the mid-stage of decomposition under highly variable moisture conditions coincide with stress-tolerant fungi dominating in the late-stage of decomposition where moisture conditions vary as a direct result of decomposition  \cite{untangling}.
    \item Though we only investigate the model under conditions where the rate of increase in dead vegetation keeps the system size $<1$m, the model represents a system of arbitrary area. Changing $l$ merely scales, and does not fundamentally change, the results.
    \item By holding temperature constant in each region, the effects of moisture on decomposition are more directly observed.
\end{itemize}

\paragraph{Limitations}
\begin{itemize}
    \item Spatial factors, such as \emph{where} the fungi and dead vegetation are, are abstracted away, though the Agent-Based Model covers this aspect.
    \item Fungi are forbidden from dying out, which prevents the model from capturing extinction.
    \item The transitive ranking system allows a single isolate to dominate under some conditions. In reality, fungal competition is intransitive, and more overall biodiversity would thus be observed in the field.
\end{itemize}

\subsection{Agent-Based Model}
\begin{figure}[!ht]
    \centering
    \begin{minipage}{0.48\textwidth}
        \centering
        \includegraphics[width=1\textwidth]{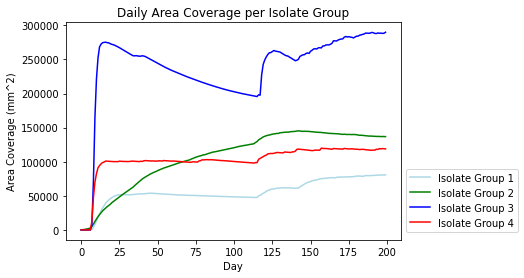}
        \caption{Daily fungal area in an arid region. \\$p$ set to .05. Total decomposition: 353,748}
        \label{fig:compare1}
    \end{minipage}\hfill
    \begin{minipage}{0.48\textwidth}
        \includegraphics[width=1\textwidth]{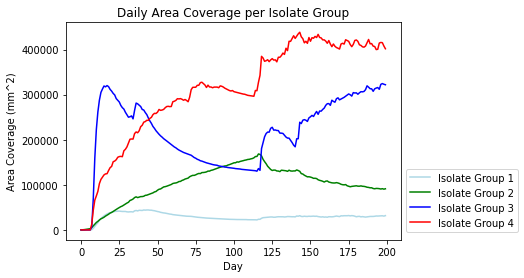}
        \caption{Daily fungal area in an arid region. \\$p$ set to 1. Total decomposition: 499,058 }
        \label{fig:compare2} 
    \end{minipage}
    \begin{minipage}{0.48\textwidth}
        \includegraphics[width=1\textwidth]{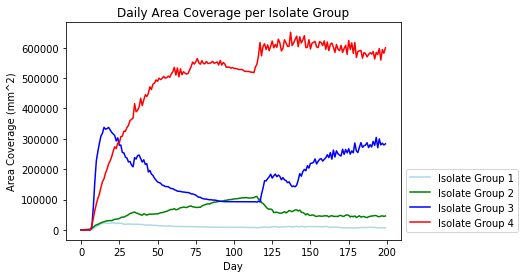}
        \caption{Daily fungal area in an arid region.\\ $p$ set to 5. Total decomposition: 648,898}
        \label{fig:compare3} 
    \end{minipage}\hfill
    \begin{minipage}{0.48\textwidth}
        \includegraphics[width=1\textwidth]{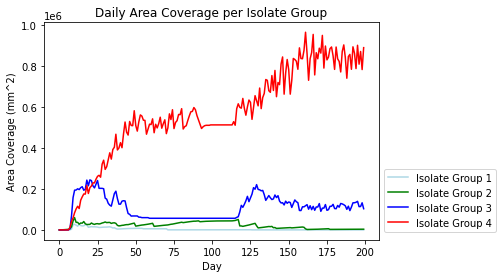}
        \caption{Daily fungal area in an arid region. \\$p$ set to 50. Total decomposition: 998,223}
        \label{fig:compare4} 
    \end{minipage}
\end{figure}

\paragraph{Parameter analysis}
We run a parameter analysis on $p$, the punishment constant, in order to better understand the impact that relative competitive capabilities have on fungi and their environments. Figure \ref{fig:compare1} demonstrates how all but nullifying group 4's innate competitive superiority in arid regions leads to a radically smaller area of coverage. In figures \ref{fig:compare2} and \ref{fig:compare3} we see results closer to those that we would expect. Group 4 out-competes the other groups in both cases. Finally, figure \ref{fig:compare4} represents the case in which innate competitive abilities of fungi are amplified to an absurd degree. As expected, group 4 totally dominates the fungi belonging to other isolates.   

We also notice that total environmental decomposition increases as group 4 becomes more dominant. This is because higher $p$ values allow the elements of group 4 to grow to larger sizes, at which point the generally faster decomposition rates exhibited by group 4 isolates will have more of an effect.  

\newpage
\paragraph{Strengths}
\begin{itemize}
    \item The model covers spatial aspects of fungal competition, which were abstracted away in the Differential Equation Model. The incorporation of space makes it more difficult for one isolate to completely dominate the system.
    \item Splitting the isolates into broad groups makes the transitivity of the ranking system less important, as general trends between groups are examined as opposed to interactions between particular isolates.
    \item Using real data for temperature allows the model to more accurately capture decomposition rates.
    \item The model considers fungi as individuals, and allows them to die, hence capturing extinction and the numbers of individuals.
\end{itemize}
\paragraph{Limitations}
\begin{itemize}
    \item Splitting the isolates into groups, while having some benefits, prevents us from discussing which particular isolates perform well under different conditions. This is, however, thoroughly covered by the Differential Equation Model.
\end{itemize}

\section{Conclusion}

By combining a Differential Equation Model which abstracts away the spatial aspects of fungal competition with an Agent-Based Model which focuses on those very aspects, we predict various changes in fungal populations and their rate of decomposition as a result of climate change.

Using the Differential Equation Model, we examine the distribution of fungal isolates in a variety of climates. By computing a metric of biodiversity on these distributions and comparing it with the variation in moisture in each climate, we see that biodiversity increases with increasing variation. This is a result of the niche differentiation of the isolates: when isolates with different moisture niches exist in a system where both niches are fulfilled from time to time, both may coexist. Moreover, the isolates that are highly competitive generally do not have wide niches, hence being drastically affected by the climate shocks and allowing other isolates to flourish when these occur. No clear-cut relationship is found between biodiversity and decomposition, and we believe more research into this area is required. The model also illustrates that a different, broader set of more stress-tolerant isolates is dominant in arid climates, when compared to the small set of highly-competitive isolates dominant in other climates. 

The Agent-Based Model corroborates these findings, examining the effects of climate on fungal interactions and overall biodiversity as they pertain to isolate groupings with similar moisture niches and growth rates.  The relative hierarchies of the different isolate groupings varied between climate regions, fluctuating as the effect of moisture conditions on competitive advantage caused fungi with decreased growth rates to be able to shift the dominance hierarchy in their favor.  As climate shocks occurred within regions, an increase in the relative proportions of fungi with higher moisture niche widths in overall populations was observed, with typically dominant fungi possessing higher growth rates but lower niche widths being unable to effectively continue to compete and maintain their wide areas of coverage.  Thus, these climate shocks, particularly in the direction of decreased moisture and increased temperatures, tended to increase overall biodiversity.   

Our parameter analyses in particular highlight the importance of field experiments in understanding how fungal interactions will impact regional decomposition rates. We advocate for continued experimental research on fungal interactions so that the factors governing their outcomes will be better understood. Specifically, the competitive rankings of fungi in real environments should be measured to determine the exact extent and rate at which competition occurs.

Global warming, in increasing desertification and average temperatures, portends a fundamental shift towards stress-tolerant fungi, accompanied by an increase in fungal biodiversity and their rate of decomposition. In increasing climate variability, global warming will further augment fungal biodiversity and the variability of their total rate of decomposition.

%%%%%%%%%%%%%%%%%%%%%%%%%%%%%%

%%% Bibliography
\addcontentsline{toc}{section}{References}
\bibliographystyle{amsplain}
\small
\bibliography{main} % filename (without the bib extension) of your bibliography file.

\newpage
\appendix 
\section{Appendix: Fungal isolates}

\begin{table}[!ht]\centering
\caption{Various properties of the fungal isolates included in the models. Note that we include multiple isolates of certain species, in which case we assign either an \emph{N} or \emph{S} to signify the Northern or Southern variant of the species, or we number the isolates arbitrarily. $R$ is the competitive rank of the isolates, $m_{opt}$ is their optimal moisture (-MPa), $w$ is their moisture niche width (MPa), and $h$ is their hyphal extension rate (m day$^{-1}$) \cite{consistent_tradeoffs}. $d$ is their decomposition rate (proportion $m^{-2}$ day$^{-1}$) at a few different temperatures \cite{trait-based}. Note that, to obtain $d$, which is the daily decomposition rate, the data was scaled exponentially from data on decomposition rate per 122 days.}

%Is $d$ supposed to be in this table? If not it seems kind of random to phrase it this way

\iffalse
$d_i$ is exponentially scaled from $D_i$ (which measures percent decomposition per 122 days) in Table \ref{table:fungi} by the function:
\begin{equation}
d_i=1-\sqrt[\leftroot{-2}\uproot{2}122]{1-\frac{D_i}{100}}
\end{equation}\fi

\label{table:fungi}
\small
\begin{tabular}{>{\itshape}lcccccccc}\toprule
& & & & \multicolumn{3}{c}{$d\times 10^{4}$}
\\\cmidrule(lr){5-7}
\normalfont \bfseries Isolate & \bfseries $R$ & $m_{opt}$ & $w$ & 10°C & 16°C & 22°C & $h\times10^3$& Group \\\hline% specify table head
\begin{filecontents*}{fungi.csv}
isolate,rank,nichemoisture,optimalmoisture,decomplow,decompmid,decomphigh,hyphae,decomplowday,decompmidday,decomphighday,nichetemperature,optimaltemperature,SpeciesGroup
Armillaria gallica 1,0.23,3.01,0.89,4.07,10.21,17.12,0.25,3.41,8.82,15.38,24.7,25,2
Armillaria gallica 2,0.16,2.08,0.78,3.2,1.89,15.42,0.35,2.67,1.56,13.72,13.3,29.8,2
Armillaria gallica 3,0.23,3.71,0.92,2.94,6.09,11,0.21,2.45,5.15,9.55,24.4,25.1,2
Armillaria gallica 4,0.23,4.3,0.85,3.78,7.23,12.3,0.25,3.16,6.15,10.75,24.3,25.8,2
Armillaria gallica 5,0.37,2.57,0.9,2.35,5.9,9.2,0.25,1.95,4.98,7.91,18.2,23.9,2
Armillaria gallica 6,0,4.15,0.68,2.03,0.11,39.51,0.4,1.68,0.09,41.12,16.2,25.7,2
Armillaria gallica 7,0.16,2.36,0.84,2.29,1.92,9.26,0.25,1.90,1.59,7.96,15.4,29.9,2
Armillaria gallica 8,0.4,2.04,0.41,2.18,0.34,10.78,0.76,1.81,0.28,9.35,9.3,26.2,2
Armillaria sinapina,0.37,1.74,0.36,1.72,6.76,8.28,0.77,1.42,5.74,7.08,16,26.4,2
Armillaria tabescens N,0.37,3.43,0.62,3.56,0.14,13.28,0.5,2.97,0.11,11.67,18.2,27.6,2
Armillaria tabescens S,0,1.32,0.27,2.83,3.67,12.75,1.07,2.35,3.06,11.17,15.8,24.7,1
Fomes fomentarius,0.28,1.19,0.24,10.41,21.26,47.24,4.71,9.01,19.57,52.27,9.3,27.3,3
Hyphodontia crustosa,0.57,1.19,0.23,4.29,13.38,13.62,0.93,3.59,11.77,11.99,23.2,25.6,1
Hyphoderma setigerum N,0.75,1.19,0.26,5.64,6.28,12.45,4.11,4.76,5.31,10.89,12.5,26.8,3
Hyphoderma setigerum S,0.57,1.38,0.56,2.67,5.63,18.82,4.7,2.22,4.75,17.08,9.7,26.7,3
Laetiporus confericola,0.28,1.22,0.56,2.29,20.28,7.6,5.16,1.90,18.56,6.48,12.1,27.1,3
Lentinus crinitus,0.57,1.55,0.31,3.47,9.3,16.01,6.38,2.89,8.00,14.29,17.8,33.8,3
Myacoacia meridionalis,0.57,1.19,0.23,2.16,5.67,7.96,1.3,1.79,4.78,6.80,12.9,25.8,1
Merulius tremullosus,0.79,1.19,0.26,22.78,32.27,53.5,10.62,21.17,31.89,62.57,17.6,30.4,4
Merulius tremellosus,0.84,1.24,0.28,13.52,15.95,43.91,9.62,11.90,14.23,47.28,18.2,30.3,4
Phlebiopsis flavidoalba N,0.59,1.57,0.33,11.18,18.52,27.94,8.04,9.71,16.77,26.82,16.2,30.7,3
Phlebiopsis flavidoalba S,0.99,2.54,0.58,6.4,8.89,25.93,10.8,5.42,7.63,24.57,17,27.4,4
Phellinus gilvus,0.46,1.4,0.6,5.15,19.47,42.09,4.04,4.33,17.73,44.68,18.6,31.5,3
Phellinus hartigii,0.49,1.57,0.65,2.3,12.86,17.39,1.54,1.91,11.28,15.65,18.6,19.1,1
Porodisculus pendulus,0.46,1.24,0.64,2.61,2.43,4.36,4.06,2.17,2.02,3.65,15,26.4,3
Phellinus robiniae N,0.52,1.53,0.63,3.68,4.52,8.28,2.3,3.07,3.79,7.08,12.3,30.2,1
Phellinus robiniae S,0.23,1.48,0.62,2.22,6.12,26.29,2.14,1.84,5.18,24.97,11.8,28.7,1
Phlebia acerina N,1,1.19,0.24,11.88,24.59,16.18,8.75,10.36,23.11,14.46,17.7,27.7,4
Phlebia acerina S,0.97,1.28,0.71,5.97,21.1,73.39,8.51,5.04,19.41,107.93,18.6,26.6,4
Pycnoporus sanguineus,0.7,1.71,0.57,4.44,18.3,37.43,4.97,3.72,16.55,38.36,17.9,37.8,3
Schizophyllum commune N,0.66,2.2,0.73,3.92,2.02,12.69,4.41,3.28,1.67,11.12,19.3,32.2,2
Schizophyllum commune S,0.59,2.32,0.82,2.55,3.32,6.87,2.57,2.12,2.77,5.83,14,32.8,2
Tyromyces chioneus,0.81,1.19,0.22,5.35,16.74,29.06,3.88,4.51,15.01,28.10,14.6,30.6,3
Xylobolus subpileatus,0.49,4.96,0.88,2.17,7.45,8.55,0.77,1.80,6.34,7.32,28.5,22.2,2
\end{filecontents*}
\csvreader[head to column names]{fungi.csv}{}% use head of csv as column names
{\isolate & \rank & \optimalmoisture &\nichemoisture & \decomplowday & \decompmidday & \decomphighday & \hyphae & \SpeciesGroup\\}% specify your columns here
\\\hline
\end{tabular}

\end{table}
\end{document}